\documentstyle[12pt]{article}
\newcommand{\beq}{\begin{equation}}
\newcommand{\eeq}{\end{equation}}

\def\half{{\textstyle{1\over2}}}

\def\p1half{{\textstyle{{{p+1}\over{2}}}}}

\def\23phalf{{\textstyle{{{23-p}\over{2}}}}}

\begin{document}
\thispagestyle{empty}
\begin{titlepage}

\bigskip
\hskip 3.7in{\vbox{\baselineskip12pt
\hbox{PSU-TH-232}\hbox{hep-th/0007056}}}

\bigskip\bigskip\bigskip\bigskip
\centerline{\large\bf 
Confinement and the Short Type I$^{\prime}$ Flux Tube}
\bigskip\bigskip
\bigskip\bigskip
\centerline{\bf Shyamoli Chaudhuri
\footnote{shyamoli@phys.psu.edu}
}
\centerline{Physics Department}
\centerline{Penn State University}
\centerline{University Park, PA 16802}
\date{\today}

\bigskip\bigskip
\begin{abstract}
We show that the recent world-sheet analysis of the quantum fluctuations 
of a short flux tube in type II string theory leads to a simple and precise 
description of a pair of \lq\lq stuck" D0branes in an orientifold compactification
of the type I$^{\prime}$ string theory. The existence of a stable type I$^{\prime}$ 
flux tube of sub-string-scale length is a consequence of the confinement of 
quantized flux associated with the scalar dualized ten-form background field 
strength $*F_{10}$, evidence for a $-2$-brane in the BPS spectrum of M theory. 
Using heterotic-type I duality, we infer the existence of an M2brane of finite 
width $O({\sqrt{\alpha^{\prime}}})$ in M-theory, the strong coupling resolution 
of a spacetime singularity in the D=9 twisted and toroidally compactified 
$E_8$$\times$$E_8$ heterotic string. This phenomenon has a bosonic string 
analog in the existence of a stable short electric flux tube arising from the 
confinement of photons due to tachyon field dynamics. The appendix clarifies 
the appearance of nonperturbative states and enhanced gauge symmetry in toroidal 
compactifications of the type I$^{\prime}$ string. We account for all of the known 
disconnected components of the moduli space of theories with sixteen supercharges, 
in striking confirmation of heterotic-type I duality. 
\end{abstract}
\noindent

\end{titlepage}

\section{Introduction}
Recently, a precise world-sheet description of the quantum fluctuations of a 
short flux tube of characteristic length $O({\sqrt{\alpha^{\prime}}})$ has been 
given, both in the bosonic and the type II string theories \cite{ccn,cn}. We will 
show in this paper that this analysis leads to a simple and elegant description
of a pair of \lq\lq stuck" D0branes in an orientifold compactification of the 
type I$^{\prime}$ string theory \cite{pw,polchinskibook,schwarz}. Away from 
the orientifold planes, the type I$^{\prime}$ theory coincides with 
the massive IIA string theory \cite{romans,pdbrane,pst} compactified on the 
interval $S^1/{\rm Z}_2$. This theory has $O(16)$$\times$$O(16)$ gauge fields 
living on the worldvolume of eight pairs of D8branes located symmetrically
on orientifold planes at $X^9$$=$$0$, $\pi R_{9I}$. The I$^{\prime}$ theory 
is T$_9$-duality equivalent to the type IB string with $32$ space-filling
D9branes carrying 
Chan-Paton labels of the group $O(16)$$\times$$O(16)$. Upon inclusion of a 
Wilson line, the IB theory can be mapped by an S-duality to the $SO(32)$ 
heterotic string compactified on $S^1$ which, in turn, is T$_9$ duality 
equivalent with appropriate Wilson line to the $E_8$$\times$$E_8$ heterotic 
string compactified on $S^1$. Conversely, an S-duality transformation maps the 
massive type IIA string theory compactified on the interval $S^1/{\rm Z}_2$, 
to the eleven-dimensional M theory compactified on $S^1$$\times$$S^1/{\rm Z}_2$. 
Following a relabeling of the coordinates $X^9$, $X^{10}$, this theory coincides 
with the strong coupling limit of the $E_8$$\times$$E_8$ heterotic string 
compactified on $S^1$. In summary, we have \cite{pw,polchinskibook}:
\begin{equation}
{\begin{array}{c}  
{\rm Heterotic } 
\\ {\rm on ~ S_1} \\ \end{array}} 
\quad 
{\begin{array}{c} S \\ \longmapsto \\ \end{array}} 
\quad
{\begin{array}{c} 
{\rm type ~ I} 
\\ {\rm on ~ S^1} \\ \end{array}} 
\quad 
{\begin{array}{c} T_9 \\ \longmapsto \\ \end{array}} 
\quad
{\begin{array}{c} 
{\rm type ~ I^{\prime}}
 \\  {\rm on ~ S^1 } 
\\ \end{array}} 
\quad 
{\begin{array}{c} S \\ \longmapsto \\ \end{array}} 
\quad
{\begin{array}{c}  
{\rm M ~ theory} 
\\ {\rm on ~ S^{1}\times S^{1}/Z_2} \\ \end{array}} 
\quad .
\label{eq:seqone}
\end{equation}
Consider a type IB background with a pair of D1branes wrapped on the compact 
$X^9$ direction, in addition to $32$ D9branes with $O(16)$$\times$$O(16)$ 
Chan-Paton labels. Let us briefly review the appearance of nonperturbative states 
in the spinor representations of $O(16)$ \cite{pw,schwarz}. A more extensive 
discussion of enhanced gauge symmetry in toroidal compactifications 
of the type I$^{\prime}$ string theory, and the appearance of disconnected 
components of the moduli space of theories
with sixteen supercharges, is given in the appendix. Although there are
by now several algebraic and/or geometric descriptions of the $E_8$$\times$$E_8$
enhanced symmetry point in nine dimensions, including bound state analyses of 
D0branes, we find the following explanation of the enhanced gauge symmetry by far 
the simplest: a T$_9$ duality transformation maps the IB background with wrapped 
Dstrings into a I$^{\prime}$ background with a pair of D0branes threading a 
stack of eight pairs of symmetrically placed D8branes at each of two orientifold 
planes. The total \lq\lq width" of each stack 
is within a distance of $O({\sqrt{\alpha^{\prime}}})$. Note that this configuration 
preserves 
the full spacetime supersymmetry of the $O(16)$$\times$$O(16)$ theory. As is well-known
\cite{pw,pst,polchinskibook,schwarz}, there is a jump of $\pm \mu_8$ in the background
9-form gauge potential each time a D0brane crosses a D8brane, associated with the
creation of a fundamental string connecting the D0brane to the D8brane. From the 
requirement of vanishing dilaton gradients at both orientifold planes, the {\em net} 
number of positive and negative jumps must exactly cancel. A sequence of jumps
may be labeled $(\pm,\pm,\pm,\pm,\pm,\pm,\pm,\pm)$, with a corresponding gradient
cancelling sequence at the other orientifold plane. Then the total number of distinct 
sequences gives the familiar counting, ${\bf 128}$$=$${\bf 2}$$+$${\bf 56}$$+$${\bf 70}$, 
for an even number of plus signs. We obtain, of course, the identical counting 
for sequences with an odd 
number of plus signs. In the limit of coincidence, the fundamental strings threading 
the stack 
of D8branes at each orientifold plane go to zero length giving massless states. 
The additional ${\bf 128}$ massless states at each orientifold plane thus enhances 
the gauge symmetry to $E_8$$\times$$E_8$.

In the absence of an external electric field, the stuck D0brane pairs threading
the pairs of D8brane stacks are absorbed into the orientifold planes 
\cite{pw}.
The dynamics of the \lq\lq stuck" pair of D0branes at sub-string length distance
scales in the presence of a constant external electric field, ${\cal F}_{09}$, 
will be the main subject of this paper. We will show that the stuck D0brane and 
its mirror image are connected by a type I$^{\prime}$ flux tube of characteristic 
length, $R$$\le$${\sqrt{\alpha^{\prime}}}$, a T$_9$ duality transformation converting 
this to a configuration of Dstrings wrapping the $X^9$ coordinate with spatial 
separation $R$ in an orthogonal direction. 
The existence of quantized constant background $C_0$ and $B_2$ gauge potentials 
on the worldvolume of the D8branes, necessitated by stability, will become evident 
in our analysis. Consider the moduli space of the nine-dimensional $E_8$$\times$$E_8$
theory. Dbrane charge conservation has been invoked to argue that while a single D0brane 
cannot move into the bulk spacetime away from the orientifold planes, a D0brane pair 
can do so. We will show that the analysis in \cite{cn} gives a simple and elegant 
solution to the problem of quantizing the fluctuations of the \lq\lq stuck" D0branes 
in a constant external electric field, at weak type I$^{\prime}$ coupling. We find 
a short distance $r^{-9}$ force between the \lq\lq dressed" D8brane stack and its mirror 
at each orientifold plane, giving evidence for the existence of $-2$-branes coupling 
to the scalar $*F_{10}$ field strength. Using the heterotic-type I duality chain 
given above, we are led to infer also the existence of an M2brane of finite width 
stretched between the orientifold planes of M theory compactified on 
$S^1$$\times$$S^1/{\rm Z}_2$ \cite{pw}.

The stability of a type I$^{\prime}$ flux tube of sub-string scale length
will be linked to the confinement of quantized flux associated with a background 
scalar dualized ten-form field strength, $*F_{10}$. Consider the T$_9$-dualized
IB theory with a pair of Dstrings wrapping the $X^9$ direction. The necessity of 
quantized constant background $B_2$ and $C_0$ fields in this IB background can
be inferred from the identification of a corresponding soliton solution of the 
IIB theory. The IB background has the classical geometry of a pair of massive 
solitonic strings, wrapped about the compact $X^9$ direction. A solitonic solution 
of the massive N=2A supergravity \cite{romans,hull} with precisely these 
characteristics was found 
in \cite{troost}; it requires background $C_0$ and $B_2$ gauge potentials. 
Application of the SL(2,Z) U-duality symmetry of the twisted T$_9$-duality
equivalent IIB string theory implies a quantization of the mass parameter
of the massive supergravity in inverse units of the radius $R_{9b}$ \cite{hull}. 
We also infer the existence of an entire $(p,q)$ multiplet of massive solitonic 
type II strings \cite{hull}. 
The geometry of the solution in \cite{troost} describes a solitonic string
perpendicular to a D8brane, but which breaks the $SO(1,1)$ symmetry unlike
the macroscopic fundamental strings of \cite{dabh}. We consider IIB 
solitonic
strings wrapped about the compact $X^9$ direction. The type IIB theory is 
self-dual, and an S-duality transformation maps this object into a Dstring 
wrapped about the $X^9$ direction. Consider lifting a IIB background with
wrapped Dstrings into a corresponding solution of the IB string theory. 
For {\em constant} background $C_0$ and $B_2$ 
fields, this is implemented by simply modding out by orientation reversal.
A T$_9$-duality transformation maps the IB background to a I$^{\prime}$ 
background with a D0brane intersecting a D8brane in a point: a \lq\lq stuck" 
D0brane, pinned to lie on an orientifold plane of the I$^{\prime}$
theory. Embedding this soliton in a compactification of the anomaly-free 
I$^{\prime}$ theory with $8$ D8brane pairs at each orientifold plane, 
completes the picture described above.

Our analysis was motivated by analogous considerations of short electric flux
tubes in the bosonic string \cite{ccn,harvey}. A world-sheet description of 
a short electric 
flux tube in the bosonic string was given in \cite{ccn}, accomplishing
in part the long-desired goal of a consistent theoretical framework for 
the short distance behavior of the Wilson loop expectation value \cite{ks}. 
The bosonic string does not have RR gauge potentials. The analysis in \cite{ccn} 
did not address the underlying bosonic string dynamics that can give rise to a 
stable electric flux tube of short length. Remarkably, precisely this 
question has been addressed in recent studies of 
tachyon condensation in open string field theory \cite{sen,harvey}, 
giving evidence for the formation of an electric flux tube with confinement 
scale of order the fundamental string length. The open strings are 
confined to a spatial volume of order $\alpha^{\prime}$. Precisely this 
behavior was found in \cite{ccn}. The heavy quark-antiquark pair interact 
via a scale invariant $-1/r$ potential 
through single photon exchange, for spatial separations restricted to 
the distance regime: 
$2\pi \alpha^{\prime}~ {\rm tanh}^{-1}v $$ \le $$ r $$\le $$ 
\alpha^{\prime} $,
where $v$ is the relative velocity of quark and antiquark. The lower 
bound is the minimum distance that can be probed by a semi-classical 
heavy quark in this background \cite{bachas}. 
The coefficient of the potential is dimensionless, a measure of
the effective number of degrees of freedom in the critical bosonic 
string: $V(r)$$=$$-(d$$-$$2)/r$. Corrections to the static term are 
parameterized by the following background field dependent variables, 
$z$$=$$|{\bf r}_{{\rm min.}}^2/{\bf r}^2|$, ${\bf u} z/\pi$, and 
$|{\bf u}|^2$, where $u^i$$=$${\rm tanh}^{-1} 
( \alpha^{\prime} F^{0i})$.\footnote{We follow the conventions
in \cite{polchinskibook}, where the $B$ field in the world-sheet action 
is dimensionless while the Maxwell field strength $F$ is dimensionful.
This also holds for the Dirac-Born-Enfield worldvolume action. The 
precise form of the $\alpha^{\prime} F$ corrections, inclusive of
numerical factors, appears in Eq.\ (48) of \cite{ccn}. Note that
no restrictions are placed on the value of the background fields 
other than the upper bound $F^{0i}$$<$$F^{0i}_{{\rm crit.}}$ in the
electric case. We should also note that, strictly speaking, the analyses 
in \cite{bachas,ccn} are carried out for magnetic background $F^{ij}$ 
fields, since the coordinate $X^0$ is obtained by an analytic 
continuation, $X^j$$\to$$iX^0$ \cite{polchinskibook,ccn}. The result can 
be appropriately interpreted in either case.}

In \cite{cn}, this world-sheet analysis was extended to the type II 
superstring theory, but with important differences due to the imposition of
consistency conditions that both eliminate the tachyon and require the 
absence of a static force between heavy quark and antiquark. The
chosen configuration of quark loops therefore preserves only one-quarter 
of the supersymmetries of the type II theory, giving rise to a short distance 
potential that is 
qualitatively different from the result quoted above:
\begin{equation}
V_{\rm super.}({\bf r}, {\bf u}) = - {{|{\bf u}|^4}\over{|{\bf r}|^{9}}}
2^{4} \pi^{7/2} {\alpha^{\prime}}^{4}
\Gamma({{9}\over{2}}) + O(|{\bf u}|^6)
\quad .
\label{eq:super}
\end{equation}
The leading term
in the potential is now {\em dimensionful}, indicating that something more
than the single photon exchange described above is at work here. Note the 
absence 
in the potential of $O(|{\bf u}|^0,|{\bf u}|^2)$ terms, and any subleading 
corrections of $O(|{\bf u}|z)$. Since the tachyon has been explicitly
eliminated, the question arises as to what plays the role of the confining
tachyon field dynamics of the bosonic string? We began by noting
that the general reasoning in \cite{harvey} can nevertheless be applied here, 
but with confinement originating in the dynamics of background RR and NS 
gauge fields. This observation led us to consider the potential for the
background gauge fields of the massive IIA supergravity theory, dimensionally 
reducing to nine dimensions where it is T-duality equivalent to a twisted 
dimensional reduction of the massless IIB supergravity \cite{hull}. This theory 
is very rich and should offer an interesting arena to study the dynamics of 
massive background gauge fields. Note that the configuration of massive 
solitonic strings breaks spacetime supersymmetry, suggesting that a full 
analysis of the moduli space dynamics, including possible tachyon instabilities 
along the lines of \cite{sen,harvey}, could be of interest. However, as explained 
at the outset, the answer to our question regarding the short type I$^{\prime}$ 
flux tube turns out to be much simpler. It relies solely on the quantization 
of scalar $*F_{10}$ flux.

Nevertheless, consideration of this theory and its solitonic solutions yields 
some important insights as explained above. Following a brief introduction to
the massive IIA supergravity theory in section 2, we identify a particular 
classical soliton, originally found in \cite{troost}, 
which takes the form of a solitonic string perpendicular to the D8brane of the
massive IIA supergravity. We compactify on a circle the coordinate $X^9$,
giving a wrapped soliton string. The soliton is T$_9$-duality equivalent to a IIB 
soliton string carrying quantized $C_0$ charge. Upon S dualizing the wrapped 
soliton string in the self-dual IIB theory, we obtain a D1brane wrapped about 
the $X^9$ direction carrying quantized $C_0$ charge. The flux of the RR scalar 
is necessarily {\em confined}: there is no corresponding propagating field. Thus, 
a configuration of a pair of closely separated Dstrings with unit winding number 
in an orthogonal compact direction is inherently stable. The quantized $C_0$
charge will give rise to a quantized theta angle for the D9brane worldvolume gauge 
fields in the IB theory as described below.

A significant simplification in the analysis of the short flux tube comes about 
by lifting the 
configuration of wrapped soliton strings into the IB string theory, which 
we do in section 3, restricting at the same time to constant background fields. 
Using simple world-sheet techniques of perturbative open 
and closed string theory \cite{backg}, and the well-known 
type I-heterotic string duality \cite{pw}, we give a solution to
the problem of quantizing the fluctuations of the short type I$^{\prime}$
flux tube.  We show that the result in \cite{cn} leads to a precise computation 
of the potential between a pair of \lq\lq dressed" D8branes at short spatial
separations of order $r^2$$<<$$\alpha^{\prime}$ in the I$^{\prime}$ theory. 
Note that, as in \cite{bachas},
the presence of a background electric field, ${\cal F}_{09}$, is crucial in
enabling a probe of sub-string length distance scales, effectively resolving 
the stuck D0brane pair. We quote the systematic expansion in powers of 
$\alpha^{\prime 4} {\cal{F}}_{09}^4$ of the potential derived in \cite{cn}, 
arising in the presence of a constant external electric field.

Finally, we come to the interesting question of the strong coupling 
dual of the short type I$^{\prime}$ flux tube. An S duality transformation
maps the short type I$^{\prime}$ flux tube into a M2brane configuration in
M theory \cite{pw,polchinskibook}: a finite width membrane of characteristic
width $O({\sqrt{\alpha^{\prime}}})$, stretched between the orientifold 
planes of the eleven-dimensional theory. The S-duality leaves the Yang-Mills
gauge fields unchanged. We infer that the membrane configuration carries 
two species of quantized instanton charge. The appendix contains auxiliary
material on the appearance of nonperturbative states and enhanced gauge
symmetry in toroidal compactifications of the IB theory with wrapped Dstrings
in addition to D9branes. As an illustration of the usefulness of the
extended heterotic-IB-IIB duality chain, exploited extensively in our 
analysis above, we provide some missing details of the heterotic-type I
duality map, accounting for all of the disconnected components of the moduli 
space of the theory with sixteen supercharges. This resolves some of the
puzzles raised in \cite{bianchi}.

\section{\bf Massive String Solitons and the Type II Flux Tube}
\label{sec:type2}

We begin by adapting the arguments for the tachyon potential in \cite{harvey} 
to that of the massive background gauge fields in the type II theory. 
The D=10 massive IIA supergravity \cite{romans} has bosonic field content 
$(G,\Phi,C_3,C_1,B_2)$, where the NS-NS two-form, $B_2$, 
is massive due to the generalized Higgs mechanism, and the $C_n$ are 
RR n-form potentials. Using the field redefinitions given in section 2 
of \cite{hull}, and with the conventions of \cite{polchinskibook}\footnote
{We assume a mostly positive signature spacetime metric and normalize
the kinetic term as follows: 
$|F_p|^2$$=$${{1}\over{p!}} G^{\mu_1\nu_1} \cdots G^{\mu_p\nu_p}
F_{\mu_1 \cdots \mu_p} F_{\nu_1 \cdots \nu_p}$.
The wedge products simply denote contractions with the appropriate
epsilon tensor; there is no metric dependence. Finally, the 
gravitational couplings satisfy the 
relation $\kappa_d^2$$=$$2 \pi R_d \kappa_{d-1}^2 $.}, 
the action takes the form: 
\begin{eqnarray}
S_{IIA} =&& {{1}\over{2\kappa^2_{10}}}
\int d^{10} X {\sqrt{- G}} \left \{ e^{-2\Phi}
[ R + 4\partial^{\mu} \Phi
\partial_{\mu}\Phi - \half |H_{3}|^2 ] - \half |F_{2}|^2
- \half |{\tilde {F}}_{4}|^2 - \half M^2 \right \}
\nonumber\\
&& \quad - {{1}\over{4\kappa^2_{10}}}
\int d^{10} X \left [  B_2 \wedge dC_{3} \wedge dC_{3} + 
M dC_{3} \wedge B_{2} \wedge B_{2} + M^2 B_2 \wedge
B_2 \wedge B_2 \wedge B_2 \wedge B_2 \right ] \quad ,
\nonumber\\
\label{eq:Lag}
\end{eqnarray}
where the gauge invariant field strengths are defined:
\begin{equation} 
F= dA + M B_2 , \quad H_3 =dB_2 , \quad {\tilde F}_4 
=dC_3 - C_1\wedge F_3 + \half M B_2 \wedge B_2 \quad .
\label{eq:field}
\end{equation}
By introducing a 9-form potential, $C_9$, acting as a Lagrange multiplier 
for the mass the action may be rewritten in terms of the RR field strength 
$F_{10}$ \cite{polchinskibook}. Note that the ten-form field strength in 
the RR sector of the IIA theory can be Hodge-dualized to a scalar field 
strength $*F_{10}$.

We assume that the $X^9$ Neumann direction is compact, a circle of radius 
$R^9$. Compactifying the massive type IIA supergravity on $S^1$ yields 
the massive type II supergravity in nine dimensions with the field content:
\begin{equation}
\left ( G_{\mu\nu},\Phi,C_{\mu\nu\lambda},C_{\mu},B_{\mu\nu} \right ) 
\oplus \left ( C_{\mu\nu 9},B_{\mu 9},
{{G_{\mu 9}}\over{G_{99}}},
{\sqrt{-G_{99}}}, C_9 \right )
\quad ,
\label{eq:fiecont}
\end{equation}
where the indices $\mu$, $\nu$ now run from $0$, $\cdots$, $8$. Note that
the subscripts denote spacetime indices here. Under a target space 
T$_9$-duality transformation, the D=9 massive IIA theory is mapped 
to the twisted dimensional reduction of the massless IIB theory 
described in \cite{hull}. The field content above is therefore isomorphic 
to the set:
\begin{equation}
\left \{ G_{\mu\nu},\Phi, C_{\mu\nu\lambda},C_{\mu\nu}, 
C_{\mu},B_{\mu\nu},B_{\mu},{\bar {A}}_{\mu}, 
e^{\chi} , {\em l}_c \right \} \quad ,
\label{eq:d9fie}
\end{equation}
where the last four entries coincide with those in Eq.\ (\ref{eq:fiecont})
above; the remaining notation is self-evident. Owing to the T$_9$ equivalence
with the IIB theory which has an inherent SL(2,R) symmetry, the p-form potentials 
are found to fit naturally into $SL(2,R)$ doublets: 
${\bar{H}}_{\mu\nu\lambda}^{(i)}$, 
${\bar{F}}_{\mu\nu}^{(i)}$, 
${\bar{B}}^{(i)}_{\mu\nu}$, 
and ${\bar {A}}^{(i)}_{\mu}$, 
where the index $i$$=$$1$,$2$ distinguishes gauge fields
originating in the NS-NS and RR sectors. We will interchangeably use 
this notation when convenient. The action takes the form \cite{hull}:
\begin{eqnarray}
S_{\rm II} =&& {{1}\over{2\kappa^2_{9}}}
\int d^{9} X {\sqrt{- G}} e^{-2\Phi}
[ R + 4\partial^{\mu} \Phi \partial_{\mu}\Phi 
- | d \chi|^2 
- \half |{\bar{H}}^{(1)}_{3}|^2 
- \half e^{2\chi} |{\bar{F}}^{(2)}_{2}|^2
- \half e^{-2\chi} |dB_{1}|^2  ]
\nonumber\\
&& \quad \quad - {{1}\over{4\kappa^2_{9}}}
\int d^{9} X {\sqrt{- G}}
\left \{ e^{\chi} [ M^2 
+ |{\bar{H}}_3^{(2)}|^2
+ |{\bar{F}}_2^{(1)}|^2
+ |{\bar{F}}_{4}|^2 ] 
+ e^{-\chi} |d {\em l}_c - M B_1|^2 \right \} 
\nonumber\\
&& \quad \quad - {{1}\over{4\kappa^2_{9}}}
\int d^9 X \left [ {\bar{F}}_4 \wedge {\bar{F}}_4 \wedge d B_{1} + 
{\bar{F}}_4 \wedge {\bar{H}}^{(1)}_3 \wedge {\bar{H}}^{(2)}_3 
\right ] \quad .
\label{eq:Lag9}
\end{eqnarray}
We have made the field redefinitions \cite{hull}:
\begin{eqnarray} 
{\bar{A}}^{(1)}_1 =&& C_1 - e^{\chi} {\bar{A}}_1 , \quad {\bar{A}}_1^{(2)} 
= {\bar{A}}_1  
\nonumber\\
{\bar{B}}^{(2)}_2 =&& C_2 - C_1 \wedge B_1 + e^{\chi} {\em l}_c 
                  {\bar{A}}_1 \wedge B_1 ,
\quad {\bar{B}}_2^{(1)}= B_2 ,
\nonumber\\
{\bar{F}}^{(1)}_2 =&& d{\bar{A}}^{(1)}_1 + 
   {\em l}_c d {\bar{A}}^{(2)}_1 + M(B_2 -{\bar{A}}^{(2)}_1 \wedge B_1 )  , 
 \quad {\bar{F}}^{(2)}_2 = d {\bar{A}}^{(2)}_1 
\nonumber\\
{\bar {H}}^{(1)}_3 =&& d B_2 + {\bar{A}}^{(2)}_1 \wedge d B_1  , 
\quad {\bar{H}}^{(2)}_3 = dC_2 - {\bar{A}}^{(1)}_1 \wedge d B_1 
  - {\em l}_c dB_2 - M B_1 \wedge B_2  
\nonumber\\
{\bar{F}}_4 =&& F_4 + {\bar{A}}_1^{(i)} \wedge d {\bar{B}}_2^{(i)} 
- \epsilon^{ij} B_1 \wedge {\bar{A}}_1^{(i)} \wedge d {\bar{A}}_1^{(j)} 
+ \half M B_2 \wedge B_2 - M B_2 \wedge {\bar{A}}_1 \wedge B_1 \quad ,
\label{eq:shifts}
\end{eqnarray}
The field content and action can be mapped to that of the twisted
dimensionally reduced massless IIB theory, 
$C_{\mu\nu\lambda}$$\to$$C_{\nu\nu\lambda 9}$, $C_{\mu}$$\to$$C_{\mu 9}$,
${\em l}_c$$\to$$C_0$, with the remaining NS potentials arising from the
particular dimensional reduction and field redefinitions given in section 
5 of the first of the references in \cite{hull}. The D=9 type II theory has 
an $SL(2,Z)$ U-duality symmetry. Consider the SL(2,Z) transformation 
parameterized, $a$$=$$d$$=$$1$, $b$$=$$n$, and $c$$=$$0$. Since,
\begin{equation}
C_0 \to C_0 + n , \quad 
(p,q) \to (p-nq, q) , \quad n \in {\rm Z} \quad , 
\label{eq:qntcp}
\end{equation}
where $(p,q)$ denotes the bound state of $p$ fundamental and $q$ Dstrings,
we find that a fundamental string, $q$$=$$0$, is mapped to another fundamental 
string, but one that carries $n$ units of RR scalar charge. The significance of 
this periodicity will become apparent later on. Note that as a consequence 
of the relation between ${\em l}_c$ and $C_0$, the mass parameter of the 
massive IIA supergravity is likewise quantized in units of $R_9$ \cite{hull}: 
\begin{equation}
M = {{n \alpha^{\prime} }\over{R_{9B}}}
\quad , n \in {\rm Z} \quad . 
\label{eq:qntp}
\end{equation}
The nontrivial potential for the massive $B$ field in the massive IIA
supergravity theory is reminiscent of that for the tachyon field in the 
bosonic string. An analysis of both kinetic and potential terms in this action 
would be complicated. However, it is always possible to work in a scaling 
limit in which the potential dominates over the kinetic terms in the action. 
This is the essential observation used in \cite{harvey}, permitting 
arguments in favor of the formation of localized macroscopic closed string 
solitons.

Accepting for the moment the ansatz that a stable and localized flux 
tube can form in the type II theory, what are its consequences? We begin 
by identifying it
with a particular soliton solution of the D=9 massive IIA supergravity. 
Soliton solutions of the massive IIA supergravity theory have been studied 
in \cite{troost}. Among those with vanishing 3-form background, and
preserving also an $SO(8)$ symmetry, is the \lq\lq massive" string:
\begin{eqnarray}
B_{09} =&& -\half f(r,z) dz \wedge dt , \quad\quad f(r,z) = 1+ {{k_f}\over{r^6}}
 + M (z - {{M r^2}\over{16}}) 
\nonumber\\
e^{-2\Phi} =&& f^{-1}, \quad\quad {\em l}_c = e^{\chi} = f^{3/8} 
\nonumber\\
ds^2 =&& f^{3/4} \left ( (dz)^2 + (dt)^2 \right )
+ f^{-1/4} \left ( (dr)^2 + r^2 (d\Omega_7)^2 \right ) 
\quad ,
\label{eq:soliton}
\end{eqnarray}
where we have defined, $t$$=$$ X^0_0$, $z$$=$$X^9_0$, 
and $r$ is the radial coordinate in the 8-dimensional
space perpendicular to the massive string. It was noticed in 
\cite{troost} that the spacetime metric of the massive string soliton 
reduces to that of 
the macroscopic fundamental strings found in \cite{dabh}, upon setting 
the mass parameter, $M$, to zero. Note, however, that even for $M$$=$$0$,
{\em there is a nonvanishing background for the RR scalar}. 
Compactifying $X^9$ on a circle, $S^1$, of radius $R_9$, gives
a wrapped massive string soliton.

From the U-duality of the equivalent IIB theory, we can infer both the 
quantization conditions in Eqs.\ (\ref{eq:qntcp}),(\ref{eq:qntp}),
and also the existence of an entire $(p,q)$ multiplet of massive string 
solitons. The type IIB string theory is self-dual, and an S-duality 
transformation exchanges RR and NS-NS backgrounds, mapping the singly 
wound massive fundamental string to a multiplet of massive $(p,q)$ strings 
wrapped on the circle $S^1$:
\begin{equation}
(1,0)  
\quad 
{\begin{array}{c} S \\ \longmapsto \\ \end{array}} 
\quad
 (0,1) \quad
  {\begin{array}{c} C_0 \to C_0 + n \\ \longmapsto \\ \end{array}} 
   \quad (-n,1)
\quad .
\label{eq:chain}
\end{equation} 
The quantum number $n$ corresponds to instanton number. This leads to
a theta term in the action of the Yang-Mills gauge fields lying in the 
worldvolume of the D1branes, obtained in the last step of the duality 
chain. The symmetry $C_0$$\to$$C_0$$+$$n$ corresponds to the periodicity of the
theta angle, $\theta$$\to$$\theta$$+$$2n\pi$. Fixing this symmetry 
nonperturbatively restricts the soliton to carrying unit instanton
winding number, $n$$=$$1$.

Finally, consider a pair of massive solitonic strings wrapped 
about the $X^9$ direction, with possible spatial 
separation $R$ in a direction orthogonal to $X^9$. Under S-duality,
we obtain a pair of D1branes coupled to a background NS two-form 
field. Following \cite{harvey}, let us \lq\lq tune" the potential 
to the point in the moduli space 
corresponding to the IIB vacuum with {\em vanishing} D1brane 
charge.\footnote{Such a tuning is hypothetical in the framework 
of \cite{cn}, given the absence of the tachyon field, but it helps 
clarify the physical interpretation of these results.}
This finally brings us to the setup in \cite{cn}: the Wilson loops are loops of 
closed string with Dirichlet boundary conditions for all embedding spacetime 
directions, but in a IIB vacuum with vanishing energy density 
for the RR $F_3$ field. 
In the \lq\lq $n$" theta vacuum the Wilson loops carry $n$ units of 
instanton charge. What is the significance of this charge? The answer 
is a form of flux confinement, although for reasons quite different 
from those at work in the tachyon field dynamics discussed in 
\cite{harvey}. This question can be addressed much more clearly upon 
lifting the IIB solitons into a background of the type I theory, as 
we now demonstrate.

\section{\bf Flux Quantization and the Type I Flux Tube}
\label{sec:type1}

To address the question posed above, we begin by noting that the mass 
parameter of the massive IIA theory can be expressed as the expectation 
value of the IIA scalar field strength, $*F_{10}$, with trivial 
equation of motion:
\begin{equation}
d *F_{10} = 0 , \quad\quad *F_{10} = {\rm constant} \quad .
\label{eq:f10}
\end{equation}
$*F_{10}$ does not give rise to a propagating field, but can nevertheless
contribute to the vacuum energy density. Under T$_9$-duality, we have
mapped the IIA background to a background with quantized expectation
value for the IIB RR scalar, $C_0$. Open strings terminating 
on the Wilson loops 
and carrying $*F_{10}$ charge would be confined by the resulting linear 
potential.\footnote{We should note here the early observation in 
\cite{pst}, and comments following Eq.\ (12.1.23) 
in \cite{polchinskibook}.} Why is the confinement scale as small 
as $\alpha^{\prime}$? We don't have a good answer, except to point out
that this is consistent with the assumption that the end-points of open 
strings carry $*F_{10}$ charge: they would {\em have} to be confined.
Observations similar to this have appeared in many places in the Dbrane 
literature \cite{pw,polchinskibook,douglas}. To the best of our knowledge, 
this is the first 
time the notion of flux quantization as an avenue for confinement has been 
applied to macroscopic soliton strings.

A much cleaner analysis emerges if we restrict ourselves to constant 
background fields, simultaneously lifting the soliton configurations
described above 
into type I string theory. We will find that the computations become 
remarkably simple, giving concrete evidence for the validity of our 
conjecture. We will use only perturbative techniques of open and closed 
string theory in constant background fields \cite{backg}, as well as
the well-known type I-heterotic string duality \cite{pw}.
The action for the type IB string theory, dimensionally reduced to
D=9, is obtained by setting to zero the IIB fields, $C_0$, $B_2$, $C_4$,
and also their dimensional reductions, $B_1$ and ${\bar A}_1$, in
Eqs.\ (\ref{eq:Lag9}) and (\ref{eq:shifts}). This leaves us with the
simplified action:
\begin{eqnarray}
S_{\rm I} =&& {{1}\over{2\kappa^2_{9}}}
\int d^{9} X {\sqrt{- G}} e^{-2\Phi}
[ R + 4\partial^{\mu} \Phi \partial_{\mu}\Phi 
- | d \chi|^2 ] 
\nonumber\\
&& \quad - {{1}\over{4\kappa^2_9}}
\int d^{9} X {\sqrt{-G}}
 e^{\chi} [ |{\tilde{F}}_3|^2
+ |{\tilde{F}}_2|^2 ] 
- {{1}\over{2g_9^2}}
\int d^{9} X {\sqrt{-G}} e^{-2\Phi}
{\rm Tr}_V |F_2|^2
\quad ,
\label{eq:type1}
\end{eqnarray}
where $F_2$ is the generalized DBI two-form field strength on the 
world-volume of D9branes wrapped on the circle $S^1$, originating
in the open string sector. We have defined:
\begin{equation}
{\tilde{F}}_3 = dC_2 - {{\kappa_{10}^2}\over{g_{10}^2}}~ {\rm Tr}_V 
\left [ A_1 \wedge d A_1 - i {{2}\over{3}} A_1 \wedge A_1 \wedge A_1
\right ] , \quad  {\tilde{F}}_2 = d C_{\mu 9} , \quad 
F_2 = dA_1 , \quad  e^{\chi} = {\sqrt{G_{99}}} \quad ,
\label{eq:filds1}
\end{equation}
where the trace denotes the sum over the vector representation of
the nonabelian gauge group, and $A_1$ is the DBI vector potential.
The Z$_2$ moded orientifold
projection leaves open the possibility of half-integer shifts in the
constant value of the background $B$ field, and integer shifts in the
constant value of the RR scalar.\footnote{It would be interesting to 
investigate type I backgrounds in $D$$<$$10$ with quantized constant 
background 5-form field strength. We are not aware of such an analysis.}
The presence of quantized constant $C_0$ and $B_2$ background fields
on the worldvolume of the D9branes
shows up in Chern-Simons couplings to the DBI field strength:
\begin{eqnarray}
I_1 =&& \int d^{10} X  ~ C_0 ~ {\rm Tr}_V \left ( B_2 
   \wedge F_2 \wedge F_2 \wedge F_2 \wedge F_2 
\right )
\nonumber\\
I_2 =&& \int d^{10} X ~  
 {\rm Tr}_V \left ( B_2 \wedge F_2 \wedge F_2 \wedge F_2 \wedge F_2 
\right )
\quad .
\label{eq:cs}
\end{eqnarray}
There has been extensive analysis of type I orientifolds with 
half-integer background $B_{\mu\nu}$ fields \cite{bianchi}. Under
heterotic-type I duality, such orientifolds are known to map into 
variants of the CHL string with $B_2$ mapped to a quantized theta 
angle in the heterotic string theory \cite{chl}.
Consider the massive string soliton given in Eq.\ (\ref{eq:soliton}).
The solution has an asymptotically flat limit when $M$$\to$$0$ and the
background fields take the constant values, $B_{09}$$=$$\half$, 
${\em l}_c$$=$$1$, and $e^{\chi}$$=$$1$ (this is with the radius of
the $X^9$ coordinate rescaled to unity).\footnote{Type I soliton
strings
with similar solution for the metric and dilaton fields were obtained 
in \cite{dbhul}. An important difference here is the presence of the 
RR scalar charge.}
At this particular point in the moduli space of constant backgrounds, 
the orientifold projection can be carried out giving a corresponding
soliton string in the type I theory. Recall that the mass parameter
is quantized in units of the radius $R_{9B}$$=$$R_{9I}$. The IB soliton 
is stable precisely because it supports RR gauge fields, including the 
quantized $C_0$ flux. A much simpler description of the type I soliton 
with unit winding number at weak type I string coupling is obtained from
the world-sheet analysis of \cite{cn}, to which we now turn.

\subsection{\bf Quantum Fluctuations of the Short Type I$^{\prime}$ Flux Tube}
\label{eq:fluct}
 
Consider quantizing the fluctuations of a short flux tube linking the
\lq\lq stuck" D0brane with its mirror image in the D=9 I$^{\prime}$ background
with $8$ D8brane stacks at each of two orientifold planes in the presence of
a constant external electric field, ${\cal F}_{09}$. As explained above, the
worldvolume of the D8branes supports, in addition, quantized constant background
fields, $B_2$ and $C_0$. Consider the duality relations \cite{polchinskibook}:
\begin{equation}
g_{I^{\prime}} \propto g_I R_{9I}^{-1} , \quad 
R_{9I^{\prime}} \propto R_{9I}^{-1} \quad .
\label{eq:dualii}
\end{equation}
Thus, for small type I string coupling, the type I$^{\prime}$ theory
is always weakly coupled even at infinite $R_{9I}$ radius. The
effect of the background electric field is to induce a background
field dependent force between the solitons, thus resolving the 
\lq\lq stuck" D0brane pair at sub-string length distance scales 
\cite{bachas}. 
T$_9$-duality maps this to a configuration of wrapped Dstrings with
small spatial separation $R$ in the $X^8$ direction, in a constant
external electric field, ${\cal F}_{09}$, lying in the worldvolume of 
$32$ space-filling 
D9branes carrying $O(16)$$\times$$O(16)$ Chan-Paton labels, in 
addition to the quantized constant $B_2$, $C_0$ fields.
Open-closed string world sheet duality permits us to describe the 
classical configuration equivalently as the \lq\lq proper" time propagation 
of a short stretched open string of length $R$, with $X^9$ identified with
open string proper time, $\sigma^1$. The endpoints of the open string 
traverse fixed
closed loops wrapped about the $(X^8,X^9)$ spacetime cylinder. The positions
of the loops on the spacetime cylinder are {\em fixed}, because they
correspond to {\em localized} type I soliton strings. Thus, the boundary 
conditions on the world-sheet are Dirichlet in all ten embedding 
coordinates of the type I string. This defines the supersymmetric
Wilson loop boundary value problem solved in \cite{cn}, where the 
consistency conditions were chosen so as to preserve one quarter of
the spacetime supersymmetries of the underlying type II theory. In
the presence of a constant background electric field, $F_{09}$, the 
endpoints of the stretched string acquire a constant velocity component 
in the $X^9$ direction relative to each other. We begin with the 
open string boundary conditions in the presence of a 
constant background two-form field \cite{backg}:
\begin{equation}
n^a \partial_a X^9 - i(2\pi\alpha^{\prime} {\cal{F}}_{09} ) t^a \partial_a
X^0 = 0, \quad 0 \le \sigma^a \le 1, ~ a=1,2 \quad .
\label{eq:bcfield}
\end{equation}
We have included the antisymmetric
tensor field, $B_{09}$$=$$\half$, in the generalized Yang-Mills field 
strength, $2 \pi \alpha^{\prime} 
{\cal{F}}_{09}$$=$$B_{09}$$+$$2\pi\alpha^{\prime}F_{09}$. 
Imposing the Dirichlet condition on both boundaries, $\sigma_2$$=$$0$,$1$, 
gives the relation
$\Delta X^9$$=$$i(2\pi\alpha^{\prime} {\cal{F}}_{09})\Delta X^0$. 
Parameterizing the
constant velocity component, 
$v$$=$${\rm tanh}^{-1} u$$=$$2\pi \alpha^{\prime} {\cal{F}}_{09}$, and
$iX^0$ as Minkowskian time, the spatial separation of the endpoints of the
stretched string takes the form:
$r^2$$=$$R^2$$+$$v^2 \tau^2$, where $\tau$ is the zero mode of
Minkowskian time \cite{polchinskibook}.

This leads naturally 
to the formulation of the string path integral with boundaries on fixed 
curves in an embedding spacetime first considered in \cite{cohen}. Let us
briefly summarize the computation of ${\cal Z}$ as given in \cite{cn}. 
The classical action for the stretched open string of length $r$ is simply 
$S$$=$$r^2 l /4\pi\alpha^{\prime}$, where we have assumed the fiducial 
world-sheet metric $ds^2$$=$$l^2(d\sigma^1)^2$$+$$(d\sigma^2)^2$, with $l$
the intrinsic length of the world-sheet boundary. The one-loop vacuum 
amplitude with boundaries on fixed curves, ${\cal C}_i$, ${\cal C}_f$,
can be rewritten in the form of a potential \cite{bachas,polchinskibook,cn}:
\begin{eqnarray}
{\cal A} =&& -i \int_{-\infty}^{\infty} d\tau ~ V_{\rm loop} [r(\tau)] 
\nonumber\\
=&& -i \int_{-\infty}^{\infty} d\tau \int_0^l dl 
~ (2\pi^2 \alpha^{\prime}/l)^{-1/2} {\rm tanh}(u) ~ 
e^{ -r^2 l/4\pi\alpha^{\prime} }  
    {\cal Z}[l,u]  
        \quad ,
\label{eq:ampl}
\end{eqnarray}
where ${\cal Z}$ is the gauge fixed one loop string amplitude including 
fluctuations of all world-sheet degrees of freedom, and the factor in 
square brackets is the normalization for the integral over the zero 
mode, $X^0_0$. Note that the zero modes for coordinates $X^1$ through $X^9$
are absent in the path integral due to the Dirichlet boundary conditions.
Gauge fixing world-sheet reparameterizations of the cylinder metric,
including the Jacobian from boundary diffeomorphisms originally computed in 
\cite{cohen}, gives the measure for the open string modulus, $l$, obtained
in \cite{ccn}. Supersymmetrizing that result, and upon 
including quantum fluctuations of both bosonic and fermionic world-sheet 
fields, we obtain the one-loop amplitude derived in \cite{cn}:
\begin{equation}
{\cal Z}[l,u] = \half N^2 \int_0^{\infty} dl \left \{ 
  {{ [\eta({{il}\over{2}})]^{-9} }\over{i\Theta_{11}(ul/2\pi,{{il}\over{2}})}}
  \sum_{(\beta,\alpha)} C_{\alpha}^{\beta} 
  \Theta^3_{(\beta,\alpha)} (0,{{il}\over{2}})
  \Theta_{(\beta,\alpha)} (ul/2\pi,{{il}\over{2}}) \right \} 
\quad ,
\label{eq:result}
\end{equation}
where we include a factor of $N^2$ for $N$ allowed values for the Chan-Paton 
index labeling a Wilson loop wrapped about the $X^9$ coordinate.\footnote{We are 
careful to distinguish $V_{\rm loop}$ from the potential between soliton strings 
wrapped on the worldvolume of 
$n$ coincident, and space-filling, D9branes. The distinction is the 
contribution to the vacuum energy density: the expression $V_{\rm loop}$
holds in the IIB vacuum without Dbranes, except for the (confined) 
flux of the RR scalar probed at short distances, $r$$<$$\alpha^{\prime}$.
The corresponding interpretation in the IB theory is explained below.}
The labels $\beta,\alpha$$\in$$0,1$ sum over spin structure and the open
string boundary conditions on world-sheet fermions, the four possible 
choices denoting R-R, R-NS, NS-R, and NS-NS sectors of the closed string, 
from the viewpoint of the closed string tree channel. Requiring the absence 
of both the tachyon and a zero field (static) vacuum energy density implies 
the phases: 
$C_0^0$$=$$-C^1_0$$=$$-C^0_1$$=$$-C^1_1$$=$$1$ \cite{polchinskibook,cn}.

The short distance behavior of the one-loop amplitude is dominated by
the term with the massless open string states circulating around the 
loop. We will perform the integration over $l$ for this term, giving the 
vacuum
energy density inclusive of the quantum fluctuations of the massless open
string modes. The result has a systematic expansion in powers of 
${\rm tanh}^{-1}(2\pi\alpha^{\prime} {\cal{F}}_{09})$$=$$u$, for 
weak fields. It takes the precise form \cite{cn}:
\begin{eqnarray}
V(r,u) =&& - N^2 (8 \pi^2 \alpha^{\prime} )^{-1/2} \int_0^{\infty} dl ~
 e^{-r^2 l/2\pi\alpha^{\prime}}
{{l^{1/2} {\rm tanh} (u) }\over{  {\rm Sin} (ul)}}
 \left [ 12 + 4 {\rm Cos} (2ul) - 16 {\rm Cos}(ul) \right ]
\nonumber\\
=&& - N^2 (8\pi^2 \alpha^{\prime} )^{-1/2}
\int_0^{\pi/u_+} dl ~ e^{-r^2 l/2\pi\alpha^{\prime}}
l^{-1/2} {{{\rm tanh}(u)}\over{u}}
\left [ \sum_{k=1}^{\infty} C_k (ul)^{2k} +
\sum_{k=1}^{\infty} \sum_{m=1}^{\infty} C_{k,m} (ul)^{2(k+m)}
\right ] ,
\nonumber\\
\label{eq:potent}
\end{eqnarray}
where the numerical coefficients in the sum are given by:
\begin{eqnarray}
C_k =&& {{ 4 (-1)^k (2^{2k} - 4 )}\over{(2k)!}}
\nonumber \\
C_{k,m} =&& {{ 8 (-1)^k (2^{2m-1} -1)}\over{(2k)! (2m)!}} ~
 |B_{2m}| ~(2^{2k} - 4)
\quad .
\label{eq:coeff}
\end{eqnarray}
The $B_{2m}$ are the Bernoulli numbers. Note that the 
$k$$=$$1$ term vanishes
in both sums and the leading background field dependence 
of the amplitude is $O(\alpha^{\prime 4}{\cal{F}}_{09}^4)$.

Integrating over $l$ gives a systematic expansion for
the potential in powers of $u^{2}/r^{4}$. As in \cite{ccn}, we identify a
dimensionless scaling variable, $z$$=$$r^2_{\rm min.}/r^2$,
where $r_{\rm min.}^2$$=$$2 \pi \alpha^{\prime} u$. We find
that the
background field dependent corrections to the leading term
in the potential are succinctly expressed as a convergent 
power series in the single variable $z$:
\begin{eqnarray}
V(r,u) =&& - N^2 (8 \pi^2 \alpha^{\prime} )^{-1/2} {\rm tanh}(u)/u
\cdot r^{-1} (2\pi\alpha^{\prime})^{1/2} [
\sum_{k=1}^{\infty} C_k z^{2k} \gamma (2k +1/2,\pi/z)
\nonumber \\
&&\quad\quad + \sum_{k=1}^{\infty} \sum_{m=1}^{\infty} C_{k,m} z^{2(k+m)}
\gamma (2(k+m) + 1/2,\pi/z)
 ]
\quad .
\label{eq:vexpn}
\end{eqnarray}
The existence of a critical limiting value for the external
electric field sets an upper bound on the regime of validity
for this result. With the restriction $z$$<$$1$, and for 
weak fields, $u$$<<$$1$, the expression in Eq.\ (\ref{eq:vexpn}) 
becomes increasingly accurate.

Now consider lifting this result into the unoriented type IB
theory. In the T$_9$-dualized I$^{\prime}$ theory, the effect of 
the constant external electric field, $F_{09}$$=$$\partial_a A_9$, 
is to {\em shift} the locations of the stacks of D8branes off of
the orientifold planes:
\begin{equation}
W = (e^{i\theta}, e^{-i\theta}, \cdots, e^{i(\pi - \theta)},
e^{-i(\pi - \theta)}), \quad X^9(0)= \theta R_{9I^{\prime}},
\quad  X^9(1)= (\pi - \theta) R_{9I^{\prime}}, \quad 
\theta_{I} = i2\pi R_{9I^{\prime}} A^I_{9} \quad ,
\label{eq:wilso}
\end{equation}
the index $I$ labeling the D8branes. We assume the constant values:
$\theta ,\pi -\theta $, for D8brane stacks at each of two orientifold
planes. Each stack is paired with its mirror image in an orientifold 
plane, and the spatial separation of a D8brane and its image is 
assumed to be small, a distance of $O({\sqrt{\alpha^{\prime}}})$. 
The theory has $U(8)$$\times$$U(8)$ gauge symmetry on the worldvolume
of the D8branes.\footnote{If the $X^0$ coordinate is interpreted as
resulting from an analytic continuation, $X^m$$\to$$iX^0$, the quantized 
background $B_{m9}$ field can imply a further reduction of the D9brane 
gauge symmetry to a single $U(8)$ (see appendix) \cite{bianchi}.} 
Note that this configuration gives vanishing dilaton and metric
gradients in the space in between the two stacks. The result of the 
electric field has been to induce a separation in the $X^9$ direction
of a given stack of D8branes and its mirror branes, symmetrically 
about each of the orientifold planes. Since we are assuming weak fields,
the separations are small, 
$\Delta X^9$$=$$i(2\pi\alpha^{\prime} {\cal{F}}_{09})\Delta X^0$. 
Parameterizing the constant velocity component as above, we see that
the spatial separation of the D0brane pair is
given by the expression obtained above, $r^2$$=$$R^2$$+$$v^2 \tau^2$, 
with $\tau$ the zero mode of Minkowskian time. The proper time evolution
of the short flux tube connecting the D0brane and its mirror, of spatial 
length $r$, 
gives a cylindrical world-sheet with boundaries, respectively, on the 
worldvolumes of a stack of D8branes and its mirror stack. The
orientifold plane slices the fiducial cylindrical world-sheet in half 
along the $\sigma^2$$=$$\half$ coordinate axis, where $\sigma^2$ is 
closed string \lq\lq proper" time. 
The resulting world-sheet can be understood as follows. A Mobius 
strip is a disc with boundary ${\cal B}_1$, with a crosscap sewn into its 
interior. Call the 
circle in the surface bounding the crosscap, ${\cal C}$. Now consider
cutting open an additional hole, ${\cal B}_2$, in the surface, along a circle 
circumscribed {\em inside} the crosscap region. The crosscap is effectively
removed, and the resulting surface is orientable, with two boundaries.
Thus, the world-sheet joining the closed world-lines of the D0brane pair
has the topology of a cylinder. As a consequence, the result for the 
potential given in Eq.\ (\ref{eq:potent}), with $N$$=$$8$, is unchanged for 
the type 
I$^{\prime}$ flux tube, and its physical interpretation is now clear. The 
potential corresponds to the short distance interaction of a pair of
\lq\lq dressed" D8brane stacks: spatially extended objects with $8$ noncompact 
spatial dimensions, carrying both D8brane and D0brane charge, in a 
generalized background electric field, ${\cal F}_{09}$. A quick check shows 
that the $r$ dependence is as expected, an $r^{-9}$ potential corresponding 
to objects with $*C_9$ charge in 
ten dimensions.

We close by noting that the \lq\lq dressed" D8brane has the correct 
properties to be identified with a $-2$-brane, a conjectured BPS
object in the massive D=9 supergravity (see comments in \cite{stelle}
and \cite{kraus}). Notice that, although we have eliminated the tachyon
in our framework, the identification of a D$-2$-brane as a D-instanton
sphaleron in the full configuration space of the massive IIA theory 
\cite{kraus} is not inconsistent with our analysis. 
Also, modulo numerical factors, which are anyhow suppressed in 
reference \cite{stelle},
the coefficient of the potential obtained above fits the form 
of the nonlinear brane-tension relations derived in \cite{stelle}
as a consequence of generalized RR flux quantization: 
$2^4 \pi^{7/2} \alpha^{\prime 4}$$=$$\tau_0$$\tau_{-2}$$\kappa^2$. 
Notice that the $u^4 r^{-9}$ 
dependence of the short distance potential also fits the expected 
behavior \cite{bachas} for a BPS state, identical at short and long 
distances. This is a bit mysterious to us at the moment. 

\subsection{Strong Coupling Dual of the Short Type I$^{\prime}$ Flux Tube}
\label{sec:dualh}

Finally, we come to the interesting question of the strong coupling dual 
of the short type I$^{\prime}$ flux tube. The type I$^{\prime}$ theory
compactified on $S^1$ has as its strong coupling dual the eleven 
dimensional M theory compactified on $S^1$$\times$$S^1/{\rm Z}_2$:
\begin{equation}
R_{10M} \propto g^{2/3}_{I^{\prime}} , 
  \quad R_{9M} \propto g_{I^{\prime}}^{-1/3}
R_{9I^{\prime}} \quad ,
\label{eq:duals}
\end{equation}
where we relabel the directions: $(9,10)_{I^{\prime}}$$\to$$(10,9)_M$. 
Thus, the $X^{10M}$ direction is a segment with boundaries on
orientifold planes, each carrying $8$ pairs of coincident D8branes. 
M theory on $S^1$$\times$$S^1/{\rm Z}_2$ is the strongly coupled
$E_8$$\times$$E_8$ heterotic string theory. Using the basic 
heterotic-type I duality relations \cite{pw,polchinskibook}: 
\begin{equation} 
g_{\rm I^{\prime}} \propto g^{-1/2}_{\rm h} R_{\rm 9h}^{3/2} , 
\quad R_{\rm 9I^{\prime}} \propto g^{1/2}_{\rm h} R_{\rm 9 h}^{1/2} 
\quad ,
\label{eq:hett}
\end{equation}
where the subscript $h$ refers to the $E_8$$\times$$E_8$ heterotic
string. Under heterotic-IB duality, the short type I$^{\prime}$ flux 
tube joining a \lq\lq stuck" 
D0brane with its mirror D0brane running in the $X^8$ direction is 
mapped to a heterotic soliton \lq\lq string" 
of length $R$$<<$$O({\sqrt{\alpha^{\prime}}})$ also in the 
$X^8$ direction. Note that lengths in the 
transverse directions are unchanged under the weak-strong coupling 
duality. The usual duality chain \cite{polchinskibook} then implies:
\begin{equation}  
{\begin{array}{c} 
{\rm \lq\lq F_8" ~ in} \\ {\rm Heterotic} \\ 
\end{array}} 
\quad 
{\begin{array}{c} T_9 \\ \longmapsto \\ \end{array}} 
\quad
{\begin{array}{c} 
{\rm \lq\lq F_8" ~ in} \\ {\rm Heterotic} \\ \end{array}} 
\quad 
{\begin{array}{c} S \\ \longmapsto \\ \end{array}} 
\quad
{\begin{array}{c} 
{\rm \lq\lq D_8" ~ in} \\ {\rm type ~ IB} \\ \end{array}} 
\quad
{\begin{array}{c} T_9 \\ \longmapsto \\ \end{array}} 
\quad
{\begin{array}{c} 
{\rm \lq\lq D_{89}" ~ in} \\ {\rm type ~ I^{\prime} } \\ \end{array}} 
\quad 
{\begin{array}{c} S \\ \longmapsto \\ \end{array}} 
\quad
{\begin{array}{c} 
{\rm \lq\lq M_{8,10}" ~ in} \\ {\rm M ~ theory } \\ \end{array}} 
\quad .
\label{eq:seqoneh}
\end{equation}
The short I$^{\prime}$ flux tube has been mapped to a finite width 
M2brane stretched between the orientifold planes in eleven dimensions.
Note that a heterotic \lq\lq string" of sub-string scale length is
no string at all and, from the viewpoint of the nine-dimensional
heterotic theory, is a singularity in the classical spacetime geometry. 
This singularity 
is resolved at strong coupling. From the duality chain given above, 
we infer that the singularity carries two species of quantized 
instanton charge. One of these, associated with the half-integer
quantized background $B_{09}$ field, is a quantized theta angle 
analogous to that characterizing toroidal compactifications without
vector structure \cite{chl}. The other theta angle corresponds to 
the quantized $C_0$ background, and can be identified with the 
familiar Yang-Mills instanton winding number. It would be extremely 
interesting to characterize this geometric singularity more precisely. 

\section{Conclusions}
\label{sec:conc}

We have shown that the recent computation \cite{cn} of the quantum 
fluctuations of a short flux tube in type II string 
theory leads to a simple description of the short distance potential 
between the \lq\lq dressed" D8branes of a D=9 type I$^{\prime}$
orientifold in a constant background electric field ${\cal F}_{09}$. 
Dressed D8branes carry both D0brane and D8brane charge, and their 
worldvolumes support {\em quantized} constant background $B_2$ and 
$C_0$ fields in addition to the usual nonabelian gauge fields. The 
nonabelian gauge symmetry is $U(8)$$\times$$U(8)$. In the absence of an
external electric field, this configuration describes a nonperturbative 
background of the I$^{\prime}$ theory with a pair of \lq\lq stuck" D0branes 
threading a stack of $8$ D8branes at each of two orientifold planes. 
Confinement is absolute in this case, and the D8brane stacks 
with stuck D0branes are absorbed into the orientifold planes as a 
consequence of the confining $*F_{10}$ flux \cite{pw,pst}. We have shown 
that the constant external electric field resolves the \lq\lq stuck" 
D0brane pair, enabling a probe of sub-string-length distance scales. 
The characteristic $r^{-9}$ fall-off of the short distance potential 
fits naturally with the interpretation of dressed D8branes as objects 
coupling to the dualized $*C_9$ potential. They can possibly be identified 
as $-2$-branes. The existence of $-2$-branes in the BPS spectrum of M 
theory has been conjectured previously \cite{stelle}, but to the 
best of our knowledge there is no known construction of such objects.

Using heterotic-type I$^{\prime}$ duality, we are led to infer 
also the 
existence of an M2brane of finite string-scale width, stretched 
between the orientifold planes in eleven dimensions. The finite
width M2brane allows interpretation as the strong coupling resolution 
of a spacetime singularity in the twisted and toroidally
compactified $E_8$$\times$$E_8$ heterotic string. From our knowledge
of the constant background fields of the IB soliton, we infer that the 
heterotic singularity carries two species of instanton charge, one of them
corresponding to the familiar Yang-Mills instanton winding number. 
The short type I$^{\prime}$ flux tube is a beautiful illustration 
of confinement as a consequence of flux quantization. 
An analogous phenomenon is believed to occur in the bosonic 
string theory, but confinement arises instead in the dynamics of
the tachyon field \cite{harvey,sen}. We have shown that the 
results in \cite{ccn} give a precise world-sheet computation, 
inclusive of numerical factors, of
the quantized fluctuations of the short electric flux tube in the
bosonic string. This includes the systematic corrections to the leading 
$r^{-1}$ static potential arising from single photon exchange. 
The corrections are expressed in the form of an expansion in powers of 
the dimensionless variable $\alpha^{\prime 4} {\cal F}_{09}^4$.

A T$_9$ duality maps the I$^{\prime}$ background into a nonperturbative 
background of the IB theory with wrapped Dstrings lying in the 
worldvolume of $32$ space-filling D9branes carrying $O(16)$$\times$$O(16)$ 
Chan-Paton labels and supporting, in addition, quantized constant 
background $B_2$ and $C_0$ fields. In the absence of the external
electric field, the nonabelian gauge symmetry is $E_8$$\times$$E_8$.
The extended heterotic-IB-IIB duality map leads to an identification 
with a particular soliton solution of the massive D=9 
type II supergravity \cite{romans,hull}. The solution takes the form 
of a \lq\lq massive" string perpendicular to the worldvolume of a
D8brane \cite{troost}. 
We use the SL(2,Z) U-duality symmetry of the nine-dimensional
type II theory to infer both the quantization of the constant background 
$C_0$, and the existence of an entire $(p,q)$ multiplet of massive 
type II solitonic strings. The mass parameter of the IIA supergravity
is quantized in inverse units of the IIB radius. An S-duality in the 
self-dual IIB theory, followed by an orientation projection at the point 
in the moduli space with quantized $C_0$ and $B_2$, gives a IB background 
with Dstrings wrapping the $X^9$ direction. Consequently, the I$^{\prime}$
soliton described above has no free parameters other than the constant
external electric field introduced in order to resolve the dressed D8brane 
stack and its mirror at each orientifold plane.

Our analysis is based on the nonperturbative D=9 IB theory 
with massless gauge bosons in the spinor representation of $O(16)$ 
\cite{pw,schwarz}. Consequently, we give a brief discussion in
the appendix of nonperturbative states and the appearance of
enhanced gauge symmetry in the toroidally compactified type IB 
string theory, in backgrounds with wrapped D1branes in addition 
to the D9branes. Such backgrounds preserve all of the spacetime 
supersymmetries of the ten-dimensional type I string, and are in 
one-to-one correspondence with the heterotic CHL orbifolds 
in spacetime dimensions $D$$\le$$9$ \cite{chl,cl}. We clarify
some missing details in the heterotic-type I duality map, addressing 
the puzzles raised in \cite{bianchi}. Using the extended heterotic-IB-IIB 
duality chain, we are able to account for all of the known 
disconnected components of the moduli space of theories with 
sixteen supercharges \cite{chl,cl}. 

\newpage
\noindent{\bf Acknowledgments}
\bigskip

This research is supported in part by the award of a grant by 
the National Science Foundation, NSF-PHY-97-22394, under the
auspices of the CAREER program. We would like to thank J. Harvey
for reminding us of relevant work in \cite{pst,kraus}. We also 
thank J. Harvey and H. Verlinde for interesting discussions.

\appendix \section{\bf 
Enhanced Gauge Symmetry in the Type I$^{\prime}$ theory}

In this appendix we clarify the appearance of enhanced gauge symmetry 
in toroidal compactifications of the Type I$^{\prime}$ string theory, 
filling in some of the missing details in the heterotic-type I duality 
map. As a consequence, we will account for all of the disconnected 
components of the moduli space of the theory with sixteen supercharges 
\cite{chl}, in striking confirmation of heterotic-type I duality \cite{pw}. 
An essential ingredient is played by toroidal compactifications of the 
IB theory with a Dstring pair wrapping any of the compact dimensions, 
in the presence of a half-integer quantized background $B$ field of even
rank. The backgrounds support, in addition, $32$ D9branes as necessitated 
by anomaly cancellation of the ten-dimensional theory. Such Dbrane 
configurations preserve all of the 
spacetime supersymmetries. They are in one-to-one correspondence with 
the CHL orbifolds \cite{chl,cl}, in precise accord with heterotic-type 
I duality. Our discussion will clarify some of the puzzles raised in the 
second of the references in \cite{bianchi}.

We begin with the type IB theory compactified on $S^1$, with $32$ 
space-filling D9branes carrying Chan-Paton indices of the group 
$O(16)$$\times$$O(16)$. Consider, in addition, a pair of Dstrings wrapped 
about the compact $X^9$ direction. This can equivalently be described in the 
T$_9$-duality equivalent Type I$^{\prime}$ theory, where it corresponds 
to a \lq\lq stuck" D0brane pair threading a stack of $8$ D8brane pairs at 
either of two orientifold planes at $X^9$$=$$0$, $\pi R_{9I^{\prime}}$. 
Massless gauge bosons in the spinor representations of $O(16)$ appear 
as a consequence of fundamental string creation arising from the crossing 
of D8brane stacks by stuck D0branes \cite{schwarz} or, equivalently, of 
space-filling D9branes with $O(16)$$\times$$O(16)$ Chan-Paton labels by 
wrapped D1branes, enhancing the 
gauge symmetry to $E_8$$\times$$E_8$ as described in the introduction. 
Thus, in {\em nine} spacetime dimensions 
and below, where the moduli spaces of the heterotic $E_8$$\times$$E_8$ and 
$SO(32)$ theories are known to be connected upon inclusion of Wilson lines, 
there exists a perfectly good type IB description of points in the moduli 
space with $E_8$$\times$$E_8$ enhanced gauge symmetry. This is in accordance 
with heterotic-type I duality. Note that the radius of the compact dimension, 
$X^9$, is arbitrary.

Now consider lowering the compactification radius, $R_{9I}$ \cite{pw}. 
The $E_8$$\times$$E_8$ heterotic string theory compactified on $S^1$ 
displays, in addition, an $SU(2)$ enhanced gauge symmetry at the self-dual
radius, $R_{\rm 9h}$$=$${\sqrt{\alpha^{\prime}}}$ \cite{polchinskibook}. This 
point in the moduli space cannot be directly accessed in the perturbative type 
IB description since the type I string coupling is growing strong (see 
Eq.\ (\ref{eq:hett})) \cite{pw}. However, we can always use the duality chain 
to first map the IB background into a corresponding IIB background by simply 
undoing orientation reversal. An S-duality transformation 
in the self-dual IIB theory is then used to map the IIB background with wrapped
Dstrings into a IIB background with wrapped fundamental strings:
\begin{equation}
{\begin{array}{c} {\rm F_{[m]} ~ in ~ IIB} \\ {\rm on S^1} \\ \end{array}} 
\quad 
{\begin{array}{c} S \\ \longmapsto \\ \end{array}} 
\quad
{\begin{array}{c} {\rm D1_{[m]} ~ in~ IIB } \\ {\rm on S^1} \\ \end{array}} 
\quad 
{\begin{array}{c} \Omega \\ \longmapsto \\ \end{array}} 
\quad
{\begin{array}{c} {\rm D1_{[m]} ~ in ~ IB} \\  {\rm on ~ S^1 } \\ \end{array}} 
\quad 
{\begin{array}{c} T_{m} \\ \longmapsto \\ \end{array}} 
\quad
{\begin{array}{c}  {\rm \lq\lq D0" ~ in ~ I^{\prime} } \\ 
{\rm on ~ S^{1}/Z_2} \\ \end{array}} 
\quad ,
\label{eq:seqoneeh}
\end{equation}
where the notation $F_{[m]}$ denotes a fundamental string wrapped about
the compact $X^m$ direction, and likewise for the Dstrings. \lq\lq D0" 
denotes a \lq\lq stuck" D0brane lying in the worldvolume of D8branes on 
an orientifold plane. The IIB theory with wrapped fundamental strings 
displays an enhanced gauge symmetry at the self-dual radius of the 
compact dimension, $X^m$, analogous to the appearance of enhanced gauge 
symmetry in the heterotic string. Thus, the precise 
counting of additional massless gauge bosons and their interactions is
most easily performed in the IIB background with wrapped $F$ strings, 
using an S-duality in the self-dual IIB theory, followed by the action of 
$\Omega$, to recover the IB spectrum with wrapped Dstrings. The precise 
match of the counting of massless gauge bosons associated
with enhanced gauge symmetry at the self-dual radii is striking confirmation
of the internal consistency of the extended heterotic-IB-IIB duality 
chain. The extension of this analysis to compactifications on $T^d$, with 
$d$$\le$$9$, is evident: the massless spectrum of a wrapped IIB fundamental
string, followed by an S-duality in a self-dual theory plus orientation 
projection, coincides with that of a wrapped heterotic fundamental string 
\cite{polchinskibook}. We are using here the fact 
that the orientation projection, $\Omega$, is a freely acting ${\rm Z}_2$ 
symmetry on the massless spectrum of the IIB theory in the presence of unbroken 
spacetime supersymmetries. In particular, it is freely reversible. Secondly, 
we used the self-duality of the IIB theory in order to identify the massless 
spectrum at weak and strong coupling. It would be extremely interesting to 
extend these considerations both for backgrounds with fewer spacetime 
supersymmetries, and for states in the massive spectrum of the nonperturbative 
IIB theory.

As an illustration of the usefulness of the extended heterotic-IB-IIB duality
chain, we will work out the IB equivalences of the CHL orbifolds \cite{chl,cl}:
${\rm Z}_n$ orbifolds of the heterotic string compactified on $T^d$, 
which leave invariant all sixteen supercharges of the D=10 theory.\footnote{
There exist, also, both type IIA/M theory and F-theory duals of the CHL
orbifold \cite{cl}. Note that these are {\em low energy} dualities. It is
not, of course, inconsistent to have more than one low energy dual for a 
given theory.} The simplest example, which occurs in all spacetime 
dimensions, $D$$\le$$9$, is to mod out by 
the ${\rm Z}_2$ outer automorphism symmetry exchanging the two $E_8$ 
lattices, accompanied by a half-integer shift in $p_m$, with $m$$\le$$d$ 
\cite{chl}. The result is a theory with sixteen supercharges and a moduli space 
characterized by $8$$+$$d$ abelian vector multiplets, {\em disconnected} from 
the moduli space of the parent 
theory with $16$$+$$d$ abelian vector multiplets. 
Its moduli space includes, in particular, an enhanced symmetry point with gauge 
group $E_8$.\footnote{Although we restrict our discussion to the massless
spectrum in what follows as in \cite{cl}, note that specifying the 
supersymmetry preserving automorphism of the $(16+d,d)$ lattice determines 
the full perturbative heterotic string spectrum. This will also hold for the
type I$^{\prime}$ dual, upon including nonperturbative states in the spinor
representations. The momentum 
lattice of the CHL ${\rm Z}_2$ orbifold has been worked out in \cite{miha}. 
A more detailed analysis of the Wilson 
lines appears in the second of the references in \cite{miha}.} Boosts in the 
$(16+d,d)$ momentum lattice, more precisely described 
as the inclusion of Wilson lines accompanied by ${\rm Z}_2$ order shifts 
in the $(d,d)$ 
compactification lattice, can give rise to many distinct enhanced symmetry points. 
These cover all of the possible simple Lie groups, and including both simply-laced 
and non-simply-laced examples \cite{chl}. More generally, modding out by any finite 
order element in the known classification of supersymmetry preserving automorphisms 
of the given $(16+d,d)$ self-dual lattices, permits a classification of the 
disconnected components of the moduli space of the theory with sixteen supercharges 
\cite{cl}. To the best of our knowledge, the explicit classification is as yet 
incomplete, although provided in principle by elements in the Monster group 
\cite{cl}. Nevertheless, many interesting corners of this problem have been 
elucidated by different authors \cite{dijtalk}. Our goal here is 
simply to identify the equivalent orbifold action on the 
I$^{\prime}$-IB-IIB theory.

Let us return to the I$^{\prime}$ description of the D=9 $E_8$$\times$$E_8$ theory 
by a pair of Dstrings wrapped on $X^9$. The T$_9$-duality equivalent I$^{\prime}$
description has a pair of stuck D0branes crossing stacks of D8branes at either 
end-point of the interval. It is evident that this theory has a ${\rm Z}_2$ 
automorphism symmetry, ${\cal I}$, under exchange of the D8brane stacks. 
Consider a half-integer shift in the $(1,1)$ compactification lattice of the 
associated self-dual D=9 IIB theory. In the presence of a half-integer quantized 
background $B$ field, the contributed shift to the momentum lattice is invariant 
under a subsequent orientation projection, surviving in the IB theory 
\cite{bianchi,polchinskibook}:
\begin{eqnarray}
p_{mL} =&& {{n_m}\over{R_{mB}}} + {{w^9 R_{mB}}\over{\alpha^{\prime}}}
  \left (G_{9m} + B_{9m} \right )  
\nonumber\\
p_{mR} =&& {{n_m}\over{R_{mB}}} + {{w^9 R_{mB}}\over{\alpha^{\prime}}}
  \left (- G_{9m} + B_{9m} \right )  
\label{eq:shiftsmo}
\end{eqnarray}
Note that the compactification radius of the circle, $R_{mB}$, coincides with
the type I radius, $R_{mI}$, invariant under the orientation projection. The
contribution from the winding modes is absent in the IB theory, but a constant
$B_{m9}$$=$$\half \alpha^{\prime}/R_{mB}^2$ gives a surviving contribution to
the $m$ component of the IB momentum, $p_m$$=$$\half (p_{mL}$$+$$p_{mR})$, 
$m$$=$$1$, $\cdots$, $d$. In addition, the IB theory has D9brane gauge fields,
and the possibility of constant background gauge fields induces a further
shift in the IB momentum from Wilson lines \cite{polchinskibook}:
\begin{equation}
p_{m} \to (n_m+ \half ) {{1}\over{R_{mB}}} 
- q^I A^I_m - {{R_{mB}}\over{2}} A_9^I A_m^I   , \quad w_9=1 \quad ,
\label{eq:shiftI}
\end{equation} 
the index $I$ labeling the D9branes. We comment that the expressions in
Eqns.\ (\ref{eq:shiftsmo}), (\ref{eq:shiftI}), are closely analogous to 
the parameterization of the $(d,d)$ momentum lattice describing a toroidal 
compactification of the heterotic string with constant background gauge
fields \cite{nsw,polchinskibook}, modulo the transformation of the radii 
\cite{pw}, and the above-mentioned quantization of the $\pm G$$+$$B$ term. 
Consider modding out by the symmetry under exchange of the D8brane stacks, 
accompanied by a translation in the compactified momentum lattice. We can 
describe this operation using orbifold terminology \cite{polchinskibook}. 
Denoting a translation vector in the momentum lattice by $v^m$, and the 
interchange of the D8brane stacks by the symbol ${\bf \gamma}$, a generic 
element of the orbifold group, $\cal H$, is denoted $({\bf v},\gamma({\bf v}))$. 
In this example, both generators are order two, giving sectors, 
$(1,{\bf v}, \gamma({\bf v}))$. The presence of the translation implies
that all of the states in the twisted sectors are heavy. Restricting to the 
untwisted sector containing states invariant under $\cal H$, the counting of 
massless gauge bosons proceeds precisely as in the CHL orbifold \cite{chl,cl}. 
The gauge bosons of $O(16)$ arise from the symmetric linear combination of
fundamental strings stretched between corresponding pairs of D8branes within 
each of two stacks. Labeling the $16$ D8branes, $(1, \cdots,8,1', \cdots , 8')$,
the untwisted massless gauge bosons correspond to fundamental string states
of zero length that are denoted:
\begin{equation}
{{1}\over{{\sqrt{2}}}} \left \{ ( [0, \cdots , \pm 1 , 0, \cdots, 0 , \pm 1 , 0, \cdots , 0 ], 0^8) + 
( 0^8 , [0, \cdots , \pm 1 , 0, \cdots , 0, \pm 1 , 0, \cdots 0]) \right \} 
\quad  , 
\label{eq:strgs}
\end{equation}
where a pair of $\pm$ signs in the $(i,j)$th locations denotes a string joining the
$i$th and $j$th D8branes, and the signs denote their orientation. The multiplicity
of $4$ for strings linking a pair of D8branes located at the orientifold plane 
accounts for the $SO(8)$ symmetry. In addition, there are massless states in the 
Cartan subalgebra corresponding to strings with both end-points on the same D8brane
denoted: $ ([0, \cdots , \pm 1 , 0, \cdots , 0],0^8) + (0^8,[0, \cdots , 
\pm 1 , 0, \cdots , 0]) $. Note that the choice of a representation of the 
fundamental strings joining D8branes by root vectors was deliberate: the 
vectors above are already in the standard normalization of length $2$. 
It is well known that there is an integer shift in the value of the 
quantized constant $*F_{10}$ background, $\nu_0$, upon crossing a D8brane 
\cite{pw,pst,schwarz,kraus}, distinguishing inequivalent vacua of the massive
IIA theory on either side of the D8brane. With the given 
normalization, $\nu_0$ can be inferred to take
values, $(n+\half){{\mu_8}\over{2}}$, as has also been noted in \cite{kraus}. 
Thus, the vacuum with mass parameter set to zero, $n$$=$$0$, corresponds to 
$\nu_0$$=$$\half$, implying a non-vanishing constant background $C_0$ 
potential in the T$_9$-duality equivalent IIB vacuum. This has already
been noted for the massive string soliton of Section 2 following Eq.\ (11). 
The 
normalizations of nonperturbative type I states in the spinor representation 
of $O(16)$ are therefore determined as given below. 
We note that the $(d,d)$ compactification lattice 
appended to the $(8,8)$ dimensional gauge lattice constructed above corresponds
precisely to the construction of the heterotic string. As a result, the action 
of an element of the orbifold group on the $(16+d,d)$ Narain lattice \cite{nsw}
determines an isomorphic orbifold action on the type I$^{\prime}$ background. 
The enhancement of the gauge symmetry to $E_8$ proceeds upon including the 
symmetric linear combinations of additional length two vectors filling out the 
spinors of $O(16)$: 
$\half(\pm,\pm,\pm,\pm,\pm,\pm,\pm,\pm,0^8)$ and
$\half(0^8,\pm,\pm,\pm,\pm,\pm,\pm,\pm,\pm)$,
with an even number of $+$ signs
as described in the introduction, which denote fundamental string creation 
due to the stuck D0brane crossing a stack of $8$ D8branes. Restricting 
to the symmetric linear combination of fundamental strings created in 
each of two stacks gives a single ${\bf 128}$. 

The extension of this construction to the abelian ${\rm Z}_n$ orbifolds listed in 
the first of references in \cite{cl} proceeds by finding a point in the moduli 
space with the required ${\rm Z}_n$ automorphism symmetry, corresponding to the
interchange of $n$ identical stacks of D8branes, accompanied by an order $n$ 
translation in the momentum lattice. The required ${\bf v}$ is identified by 
adjusting the background gauge fields appropriately, the Wilson lines 
parameterizing 
the locations of the individual D8brane stacks. We will not pursue the analysis
further in this paper, although it is evident at this point that the constructions 
in \cite{cl} map isomorphically into the type I$^{\prime}$ description {\em upon 
inclusion of nonperturbative type I states}.\footnote{This is an important point. In 
the absence of the $O(2n)$ spinor representations, the equivalence between the 
heterotic and IB gauge lattices described above would not hold.} There is an important 
issue here regarding the imposition of consistency conditions on these theories. 
In \cite{chl}, this was done using the fermionic construction, a technique of 
outstanding reliability. The $(Z_2)^k$ examples of \cite{cl} with $k$$=$$1$, $\cdots$, 
$4$ correspond to moduli spaces with $d$$+$$2^4/2k$ abelian vector multiplets, and are
well-corroborated in the fermionic construction \cite{chl}. These theories are
easily matched with type I$^{\prime}$ backgrounds with half-integer quantized
$B$ field, examples of which first appeared as the rational type I unorientifolds 
in \cite{bianchi}, as is clear from the discussion above. It should be clear from 
our discussion that, as in the heterotic string, making an arbitrary distinction
between automorphisms of the gauge lattice or automorphisms of the compactification
lattice is rather artificial, upon the inclusion of Wilson lines. Thus, there exist 
well-documented examples with fewer than $d$ vector mutiplets in both the fermionic
and heterotic orbifold constructions \cite{chl,cl}, including a four dimensional 
moduli space which is pure N=4 supergravity with no additional vector multiplets
\cite{chl}. It would be interesting to study their type I$^{\prime}$ analogs. 
We comment that the construction described above can be mapped rather straightforwardly 
into the fermionic construction of \cite{chl}. This would clarify the relationship
of the consistency conditions of the fermionic construction to tadpole 
cancellation, enabling a cleaner 
analysis of the consistency conditions in type I$^{\prime}$ theories with less 
supersymmetry. This introduces many new features into the analysis, and we postpone 
further discussion to future work.

\end{document}